\documentclass[]{aa}  
\usepackage[dvips]{graphicx}
\usepackage{txfonts}
\usepackage{natbib}
\usepackage{rotating}
\bibpunct{(}{)}{;}{a}{}{,}
\newcommand{\HII}{H\,{\sc{ii}}}

\begin{document}
\title{Near-infrared spectroscopy in NGC~7538\thanks{Based on observations made with the WHT operated on the island of La Palma by the Isaac Newton Group in the Spanish Observatorio del Roque de los Muchachos of the Instituto de Astrof{\'{i}}sica de Canarias.}
}
\author{E. Puga\inst{1}
\and A. Mar{\'{i}}n-Franch\inst{2}
\and F. Najarro\inst{1}
\and A. Lenorzer\inst{2}
\and A. Herrero\inst{2}
\and J.~A. Acosta Pulido\inst{2}
\and L.~A. Chavarr{\'{i}}a\inst{3}
\and A.~Bik\inst{4}
\and D.~Figer\inst{5}
\and S.~Ram{\'{i}}rez Alegr{\'{i}}a\inst{2}
}
\offprints{E. Puga}

\institute{Centro de Astrobiolog{\'{i}}a (CSIC-INTA), Ctra. de Torrej\'on a Ajalvir  km-4, E-28850, Torrej\'on de Ardoz, Madrid, Spain\\
\email{elena@damir.iem.csic.es}
\and Instituto de Astrof{\'{i}}sica de Canarias, C/ V{\'{i}}a L{\'a}ctea s/n, E-38200, La Laguna, Spain
\and Harvard-Smithsonian Center for Astrophysics, 60 Garden Street, Cambridge, MA 02138, USA
\and Max-Planck-Institut f\"ur Astronomie, K\"onigstuhl 17, D-69117, Heidelberg, Germany
\and Rochester Institute of Technology, 54 Lomb Memorial Drive, Rochester NY 14623, USA 
}

\date{Received; accepted }

 
  \abstract
  {}
   {The characterization of the stellar population toward young high-mass star-forming regions allows to constrain fundamental cluster properties like distance and age. These are essential when using high-mass clusters as probes to conduct Galactic studies. }
   {NGC~7538 is a star-forming region  with an embedded stellar population only unearthed in the near-infrared (NIR). We present the first near-infrared  spectro-photometric study of the candidate high-mass stellar content in NGC~7538. We obtained $H$ and $K$ spectra of 21 sources with both the multi-object and long-slit modes of LIRIS at the WHT, and complement these data with sub-arcsecond $JHK_{s}$ photometry of the region using the imaging mode of the same instrument.}
   {We find a wide variety of objects within the studied stellar population of NGC~7538.  Our results discriminate between a  stellar population associated to the \HII{} region,  but not contained within its extent,  and several pockets of more recent star formation.  We report the detection of CO bandhead emission toward several sources as well as other features indicative of a young stellar nature. We infer a spectro-photometric distance  of  2.7$\pm$0.5\,kpc, an age spread in the range 0.5$-$2.2\,Myr and a total mass  $\sim$1.7$\times$10$^3$\,M$_{\sun}$ for the older population.}
  {}
   \keywords{Stars: early-type - Stars: pre-main sequence - Infrared: stars - Galaxy: open clusters and associations: general }

   \maketitle
%

\section{Introduction}

Clusters of total  mass $>$10$^3$\,M$_{\sun}$ are the natural habitat of high-mass stars. Whether motivated by the study of the  formation, evolution of feedback effects of high-mass stars themselves \citep{Bik_10,Martins_09,Puga_09} or as probes to investigate the physics and chemistry evolution of the Galaxy \citep{Messineo_09,Davies_09}, NIR spectroscopic studies have been proven  powerful to unveil and characterise the complete high-mass stellar content within massive clusters.\\    
Although new NIR surveys, such as UKIDSS and VISTA, will surely uncover new obscured massive clusters, some clusters, already known from optical observations,  harbour a hidden  high-mass  stellar component only accessible at NIR wavelengths \citep[e.g. Cyg~OB2,][]{Knoedl_00}.\\

\begin{figure*}[!ht]
\centering
\includegraphics[bb=53 34 340 340,scale=1.0]{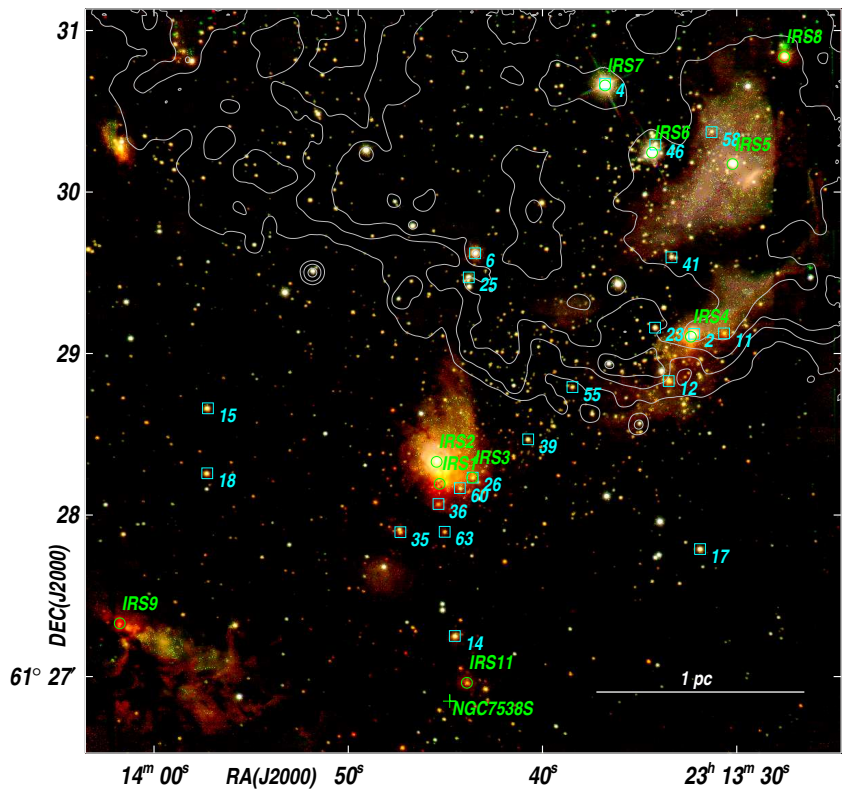}
\caption{
Colour composite image of NGC~7538 obtained with LIRIS ({\it blue}: $J$, {\it green}: $H$, {\it red}: $K_s$). Overlayed contours depict the extension of the \HII{} region in the DSS/POSSII-F Red map. Stars observed with the LIRIS--MOS mode are labelled in cyan. Object \#4 was observed with the same instrument in long-slit mode.   The green circles and cross are 2MASS point sources of interest.}\label{ngc7538}
\end{figure*}

\object{NGC~7538} (aka Sh~2-158) is a visible \HII{} region in the Perseus spiral arm and a site  of active star formation. The early detection of several luminous  NIR and far-IR sources in the vicinity of  this region \citep{Campbell_88}, hinted toward a rather massive nature. In particular,  NGC~7538 IRS~1 is a high-mass ($\sim$30\,M$_{\sun}$) protostar  with a CO outflow, known to power the  ultra-compact (UC)\HII{} region NGC~7538\,A. It also has an associated  linear methanol maser structure, which might trace a Keplerian-rotating circumstellar disk \citep{Pestalozzi_04}.
Recently, 6.7\,GHz methanol masers have been  detected  toward  the  nearby objects NGC~7538 IRS~9 and NGC~7538\,S, tracing other young and embedded massive protostars \citep{Pestalozzi}. \\
Although this region has been widely inspected at long wavelengths (mostly in the submillimetre window)   and even optical spectroscopy for two stars has been obtained \citep{Russeil_07}, only a few detailed near-IR photometric studies of the stellar population in NGC~7538 have been conducted  \citep{Balog_04,Ojha_04,McCaughrean_91}. At subarcsecond resolution, the region breaks down into several areas of interest (see  Fig.~\ref{ngc7538}): ({\sc i}) the presumed powering cluster, centred at the location of IRS~6, ({\sc ii}) the subcluster sitting amidst the bright  NIR reflection nebula (IRS~5), ({\sc iii}) the cluster located at the southern rim of the optical \HII{} region (IRS~4),  ({\sc iv}) the active region around IRS~1/2, ({\sc v}) an embedded stellar cluster at the location of NGC~7538~S, only detected at wavelengths longward of  3.6 $\mu$m, and ({\sc vi}) the IRS~9 reflection nebula.
Surprisingly, the stellar density detected in the NIR is highest at the southern rim of the dust bubble that bounds the optical \HII{} region \citep{Balog_04}, between regions ({\sc iii}) and ({\sc iv}).
The distance to NGC~7538 has been several times estimated and values between 2.2 and 3.5\,kpc have been reported \citep{Moreno_86,Israel_73}.

In this article, we present NIR spectra of a sample of candidate high-mass stars within NGG~7538 together with sub-arcsecond NIR  photometry.\\
We describe the observations and data reduction in Sect.~\ref{obs_data_red}. In Sect.~\ref{results} we analyze the spectro-photometric information of a sample of candidate high-mass stars. Sect.~\ref{discussion} is dedicated to the estimate the distance, age and mass of the cluster powering the  \HII{} region. Finally, we conclude in Sect.~\ref{conclusions} with a brief summary.

\section{Observations and data reduction}\label{obs_data_red}
\subsection{NIR photometry}\label{obs_data_red_ima}

We conducted $JHK_s$ imaging observations of the field around NGC~7538 with LIRIS \citep{Acosta,Manchado}, mounted at the Cassegrain focus of the William Herschel Telescope in La Palma. Observations were obtained  on 2006 July 21 as part of the MASGOMAS programme that comprised a total of 45 young clusters in the Milky Way (see \cite{Marin_09} for a description of the programme). The average seeing during the imaging observations of this region was 0\farcs60, 0\farcs63, and 0\farcs64 for  $J$, $H$, and $K_s$, respectively.   We followed the procedure  described in \cite{Marin_09} for the reduction and PSF extraction of the point-like sources in the field. The saturation limit for the LIRIS photometry corresponds to  $K_s\sim$ 10.5 mag. For  brighter objects, we complemented the photometry with the  2MASS point-source catalogue values that had their photometric errors determined.

\subsection{NIR spectroscopy}\label{obs_data_red_spect}

Follow-up spectroscopic observations of NGC~7538 were obtained in 2007 September 20, and 2008 August 19  using the multi-object spectroscopic (MOS) mode of the LIRIS Instrument. The average seeing during  both nights was 0\farcs9.
The designed mask comprised 20 slitlets of 0\farcs8 in width and  6\arcsec  in length. We used the low resolution grism ($\lambda$/$\Delta \lambda\sim$ 700)  that covers the $H$ and $K$ band simultaneously and the medium-resolution grisms in the $H$ and $K$ bands with $\lambda$/$\Delta \lambda\sim$ 2500 and $\lambda$/$\Delta \lambda$ $\sim$ 1700, respectively. The slitlet mask was oriented along a P.A.$=$\,142\degr\, and centred at $\alpha_{2000}$$=$\,23$^h$\,13$^m$\,39$^s$,\,
$\delta_{2000}$$=$\,$+$61\degr\,28\arcmin\,60\arcsec. \\
We obtained the observations following  a classic ABBA telescope nodding pattern with an offset of 2\farcs5. 
In the case of two slitlets, the spectra  appeared contaminated by other sources that fell close to  one of the nodding  positions (position B). Therefore,  we extracted those specific spectra on only  half of the stacked frames (at the position A). Observations of the AOV standard star HD223386 were obtained in the different configurations in order to remove the telluric features due to the Earth's atmosphere.\\ 
The data were reduced using LIRISDR\footnote{http://www.iac.es/project/LIRIS} a specific data reduction package developed under the IRAF environment by the LIRIS team. The routines available for the LIRIS MOS mode make use of the a-priori mask design information to trace the slitlets and their limits in the spatial direction in a uniformly illuminated frame. Special care was taken to make the flat-field correction. The limited spatial extent of each slitlet may include detector defects at comparable or larger spatial scales which are not detected within individual flat-field slitlets. We circumvented the problem by coadding the spectral response information of all apertures into a unique monodimensional response, after shifting by  an appropriate offset along the spectral axis. This response was
modelled using a high order spline function. The pixel-to-pixel  correction was determined for each slitlet from the extracted flat-field spectrum divided by the modelled response.
\begin{figure*}[!ht]
\begin{center}
\includegraphics[bb=29 11 326 381,scale=0.7]{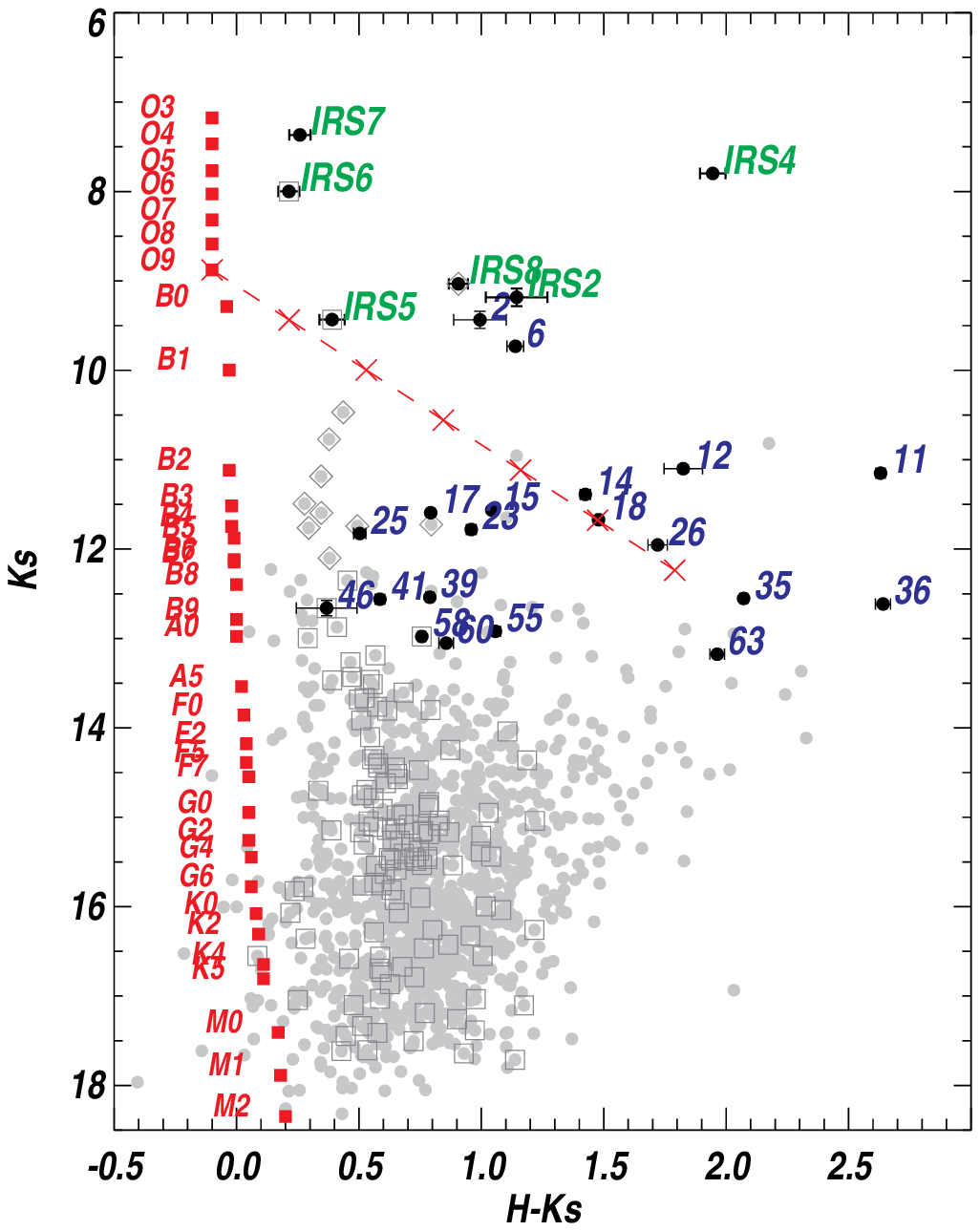}\includegraphics[bb=56 11 340 381,scale=0.7]{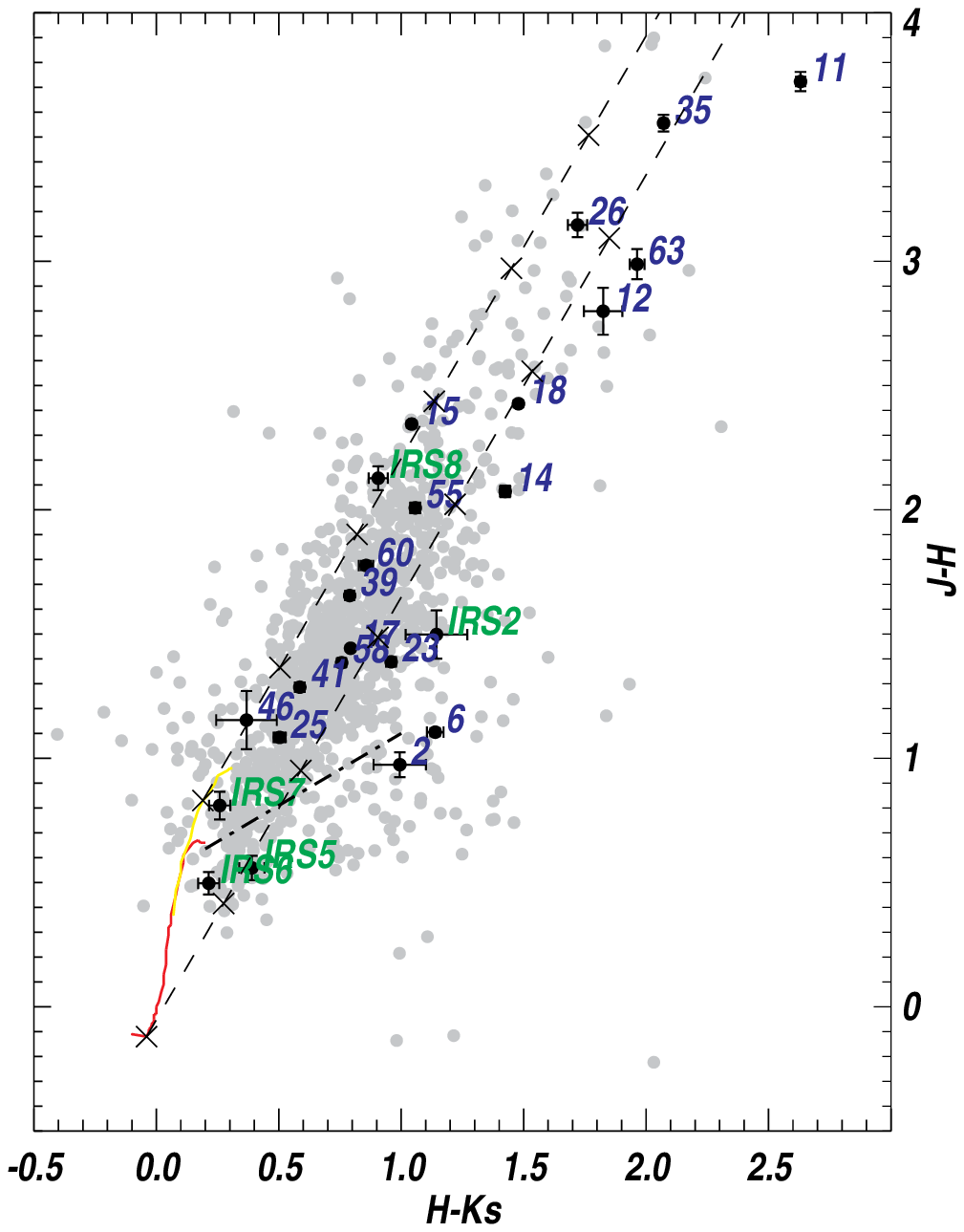}
\caption{
{\it (Left panel)} CMD of the stellar content within
the field of view of the left panel. The  red squares
represent the main sequence (MS) of dwarf stars without extinction at a heliocentric distance of 2.7\,kpc. The red  discontinuous line represents the reddening vector for an O9V star, with intervals of A$_{V}$ $=$ 5 mag indicated by crosses. Dark squares represent objects located at a projected distance smaller than 30$\arcsec$ from IRS~6. {\it (Right panel)} Colour-colour diagram of NGC~7538. Source \#36 lies at [4.84, 2.64], out of the scale of the figure. The dashed lines indicate the direction of extinction
according to \cite{Rieke&Lebofsky} with overlayed intervals of A$_{V}$ $=$ 5 magnitudes indicated by crosses. The dash-dotted line represents the T~Tauri locus of \cite{Meyer}.  
}\label{CMD_CCD}
\end{center}
\end{figure*}
We performed a standard subtraction of consecutive AB pairs to remove the sky background. Individual  flatfielding, retracing, resizing and extraction for each star in every subtracted frame  followed. Argon and Xenon arc frames  were used to determine the wavelength calibration (rms\,$<$\,1\AA).  The resulting spectra were coadded weighting by the relative number of counts to provide a final spectrum of each star. \\
The wavelength calibrated spectra were divided by a normalized reference spectrum to remove the telluric absorption features. This normalized reference spectrum was generated from the observed spectra of the A0V standard star, divided by a Vega model spectrum convolved to the correspondent spectral resolution. For this,
we constructed an IDL routine that iteratively fits the components due to the instrument response, an atmosphere's transmissivity template and the Vega model spectrum.

Source \#4 was observed with LIRIS long-slit mode in the ranges 1.92$-$2.4\,$\mu$m and 1.52$-$1.78\,$\mu$m in the nights of 2007 September 21 and 2008 June 27, respectively. The intermediate-resolution grisms provided  spectral resolutions of $\lambda$/$\Delta \lambda\sim$ 1700 and $\lambda$/$\Delta \lambda\sim$ 2500. In this case, we traced the spectrum using the continuum of the source and applied an optimal weight to extract the flux along the spatial direction. The standard star HD223386 was observed in the same configurations in order to obtain the telluric correction.\\  

\section{Results}\label{results}
\subsection{Colour-magnitude  and colour-colour diagrams}

The left and right panels of Fig.~\ref{CMD_CCD} show the colour-magnitude diagram (CMD) and the colour-colour diagram (CCD) of the field around NGC~7538.   In the figure, we have only depicted those objects that have LIRIS counterparts in the $H$ and $K_s$ bands and in the  three channels for the CMD and CCD, respectively.\\ 
The 2MASS source IRS~4 appears resolved into two components in our LIRIS images \citep[see also][]{Ojha_04}. The brighter source to the southeast is saturated in our $K_s$ map, while the northwest source \#2 is one of the brightest in the field. Therefore, we use our LIRIS photometry for source \#2, while still display the 2MASS integrated photometry of these two point-like sources in our CMD (IRS~4). IRS~4 is not detected  in the 2MASS $J$ band survey, therefore, it is important noticing that it shows $J$$-$$H$$\ge$1.34 and $H$$-$$K_{s}$$=$1.94$\pm$0.05,  pointing to a strong NIR excess.\\
This paper aims at identifying the high-mass stellar content of a Galactic massive cluster, thus we selected the candidates for spectroscopic follow-up using a NIR photometric criterion.  We looked for  reddened bright objects in high stellar density regions. The production design specifications of the slitlet mask hinder a complete study of extremely crowded regions.  
A geometrical limitation is also imposed by the LIRIS multi-object mode. The spectral coverage for each object depends on the position of the corresponding slit relative to the centre of the chip in the  dispersion direction. Therefore,  the follow-up sources must be selected in a preferred orientation along the sky.\\
\begin{figure*}[!ht]
\centering
\includegraphics[scale=1.15]{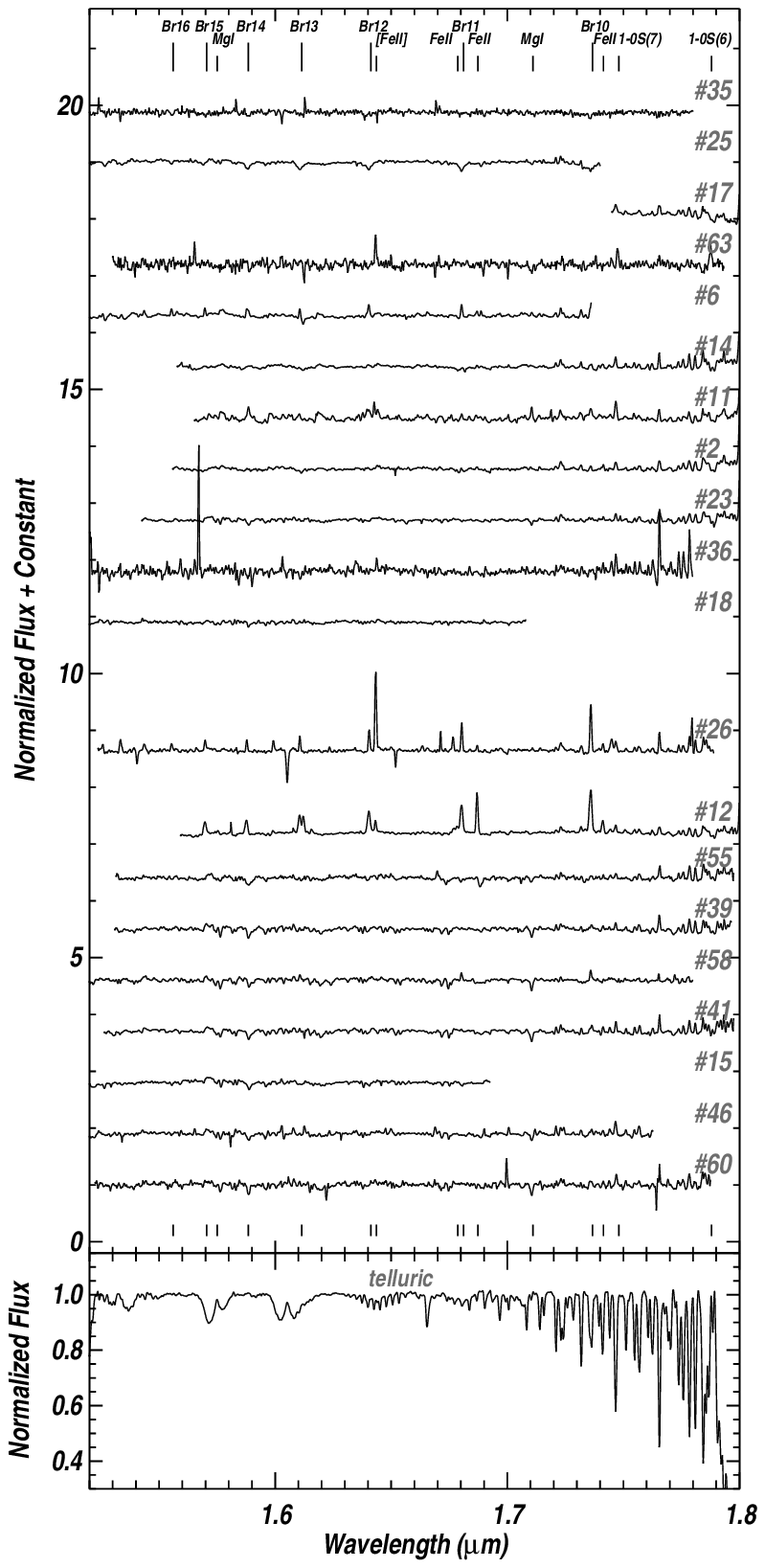}\includegraphics[scale=1.15]{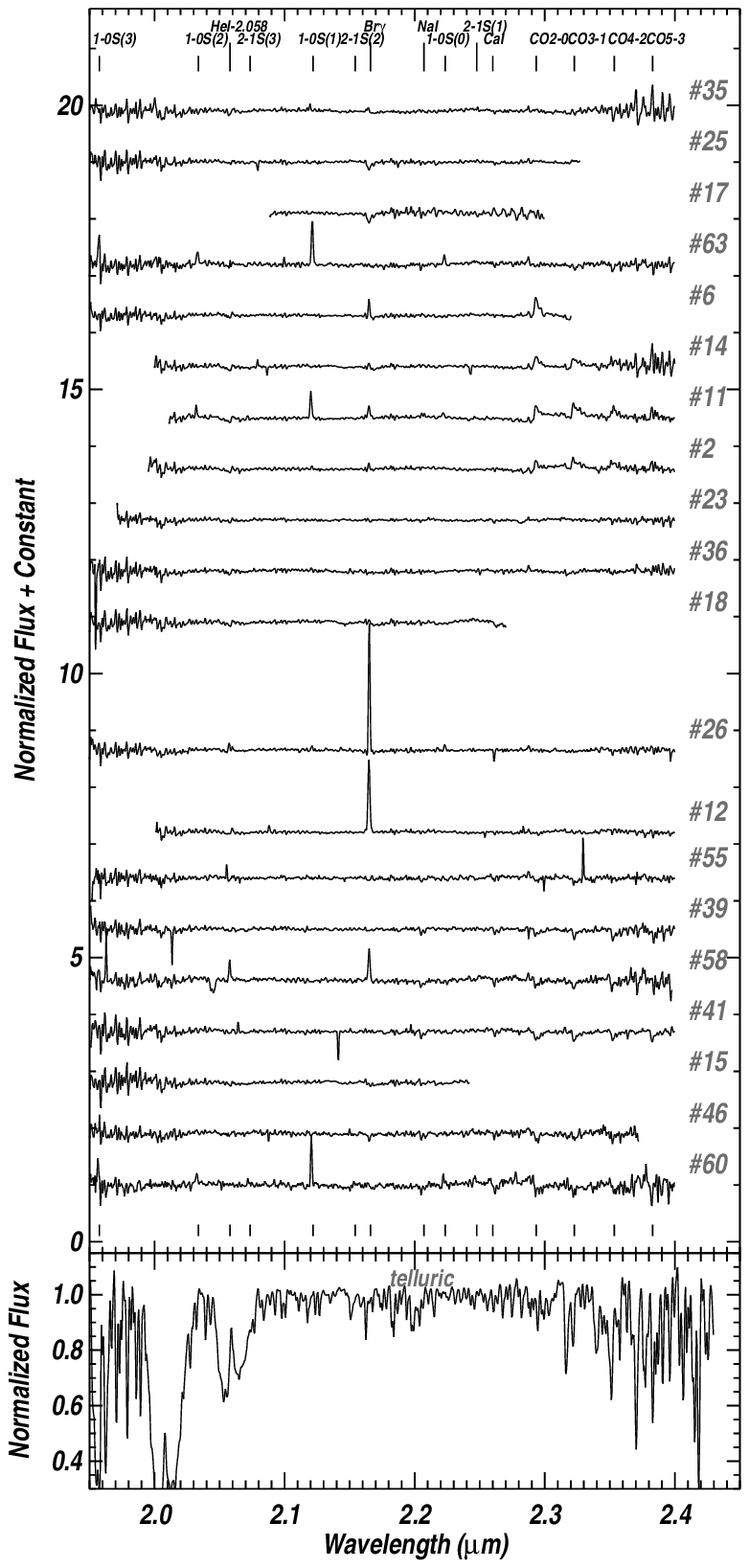}
\caption{{\it (Left panel)} H-band normalised spectra of the 20 sources with MOS (R$\sim$2500). A normalised template telluric spectrum is shown in the bottom panel. {\it (Right panel)} K-band normalised spectra of the 20 sources with MOS (R$\sim$1700). A normalised template telluric spectrum is shown in the bottom panel.}\label{fig_spectra}
\end{figure*}
Our CCD evidences the strong variations in  extinction toward the sources selected for follow-up spectroscopy.

\subsection{Spectral classification}\label{spect_class}

\begin{figure*}
\centering
\resizebox{\hsize}{!}{
\includegraphics[scale=0.9]{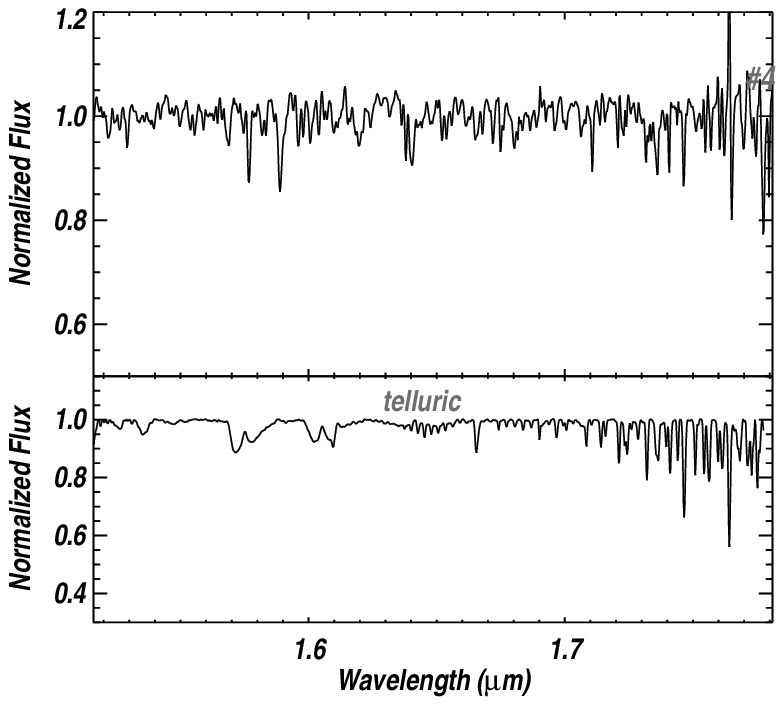}\includegraphics[scale=0.9]{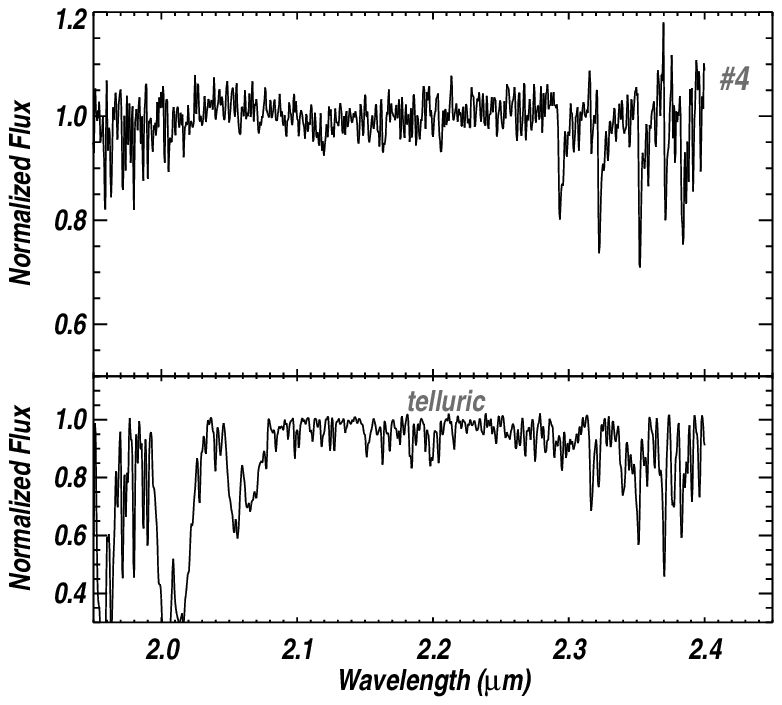}
}
\caption{{\it (Left panel)} H-band normalised spectrum of source \#4 obtained with long-slit (R$\sim$2500). {\it (Right panel)} K-band normalized spectrum of the same source with long-slit (R$\sim$1700). A normalised template telluric spectrum is shown below. }\label{fig_spectra_4}
\end{figure*}

\begin{figure*}
\centering
\includegraphics[scale=1.15]{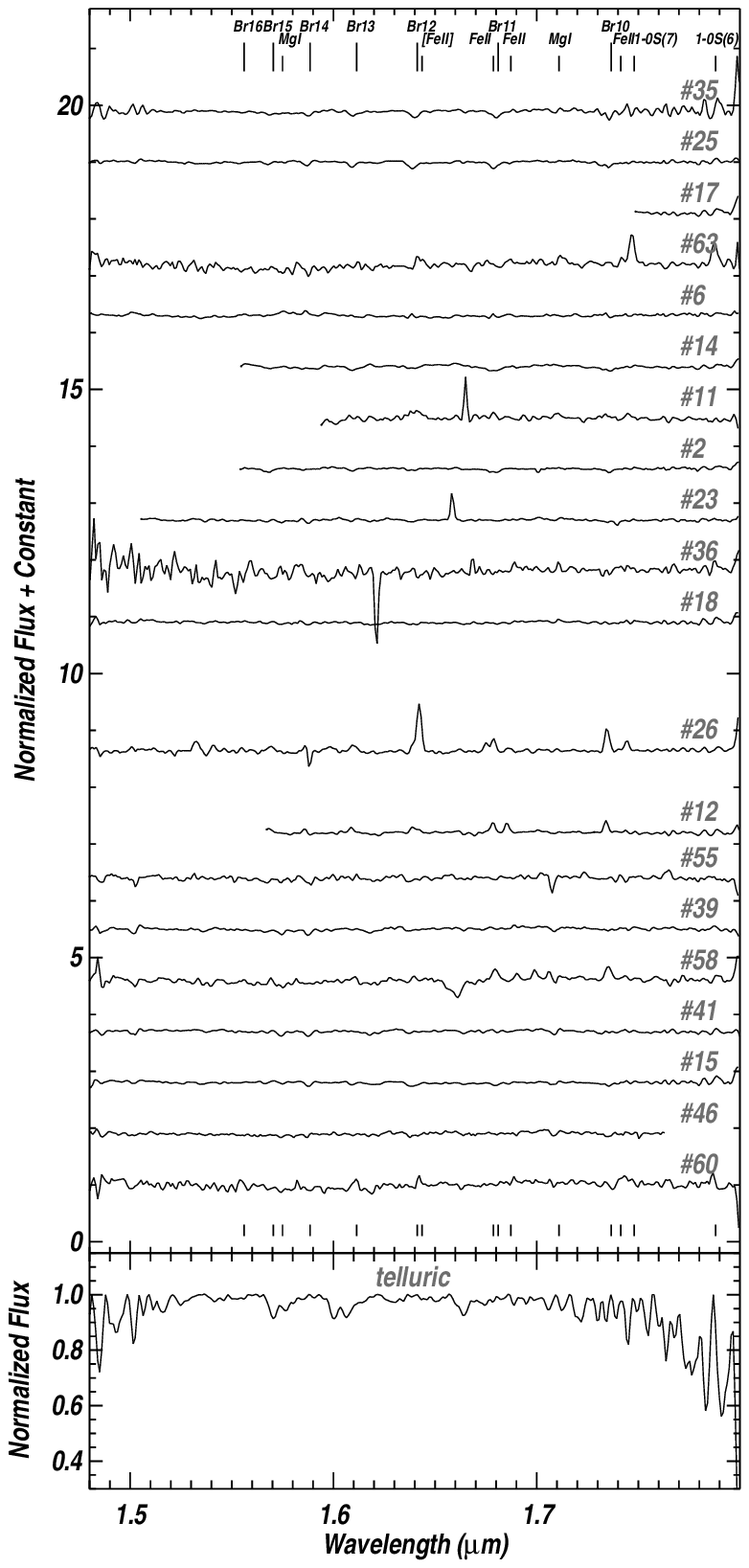}\includegraphics[scale=1.15]{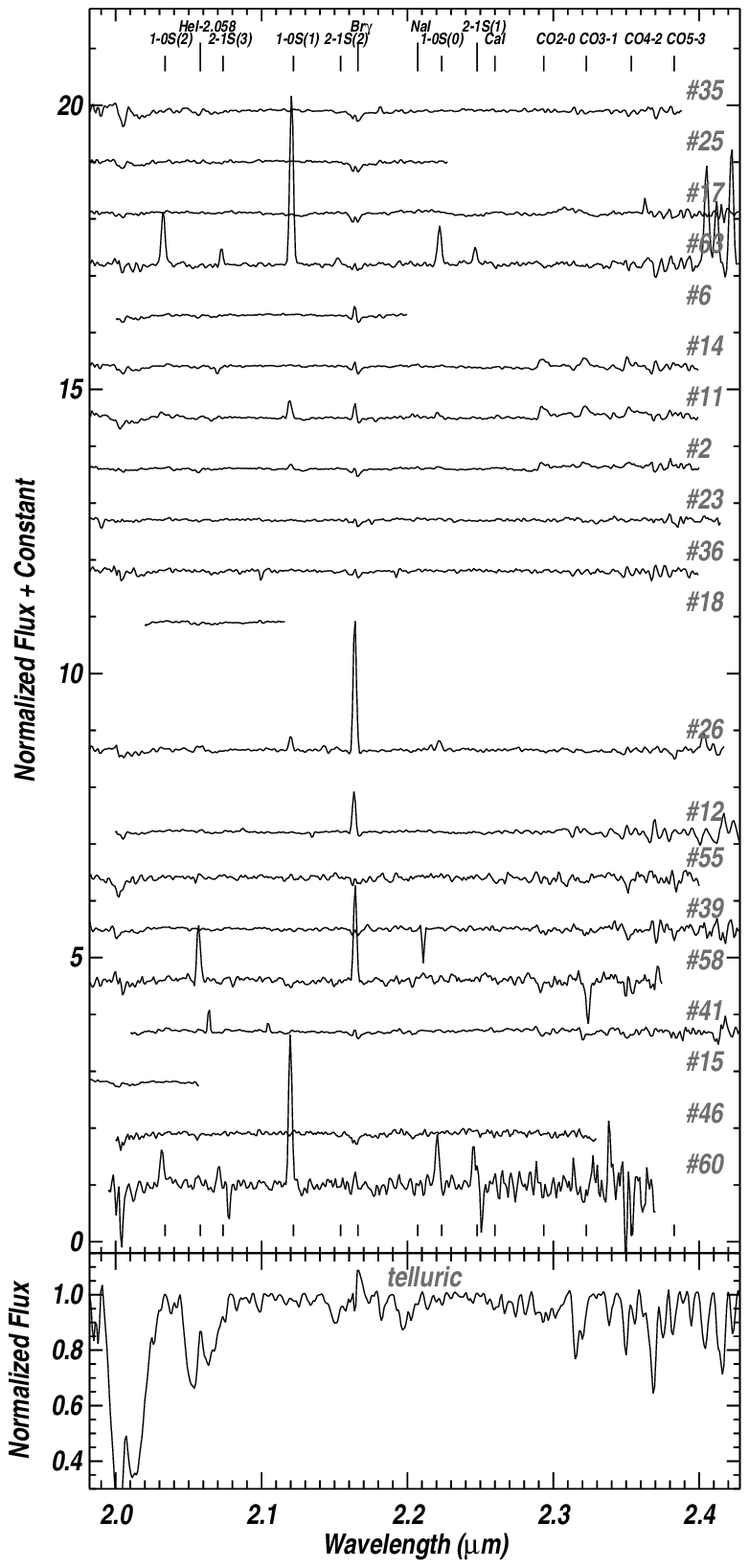}
\caption{H- and K-band normalised spectra of the 20 sources with MOS (R$\sim$700). A normalised template telluric spectrum is shown in the bottom panel. }\label{fig_spectra_lr}
\end{figure*}

\begin{figure}[!h]
\begin{center}
\resizebox{\hsize}{!}{
\includegraphics[bb=15 8 122 337, scale=0.65]{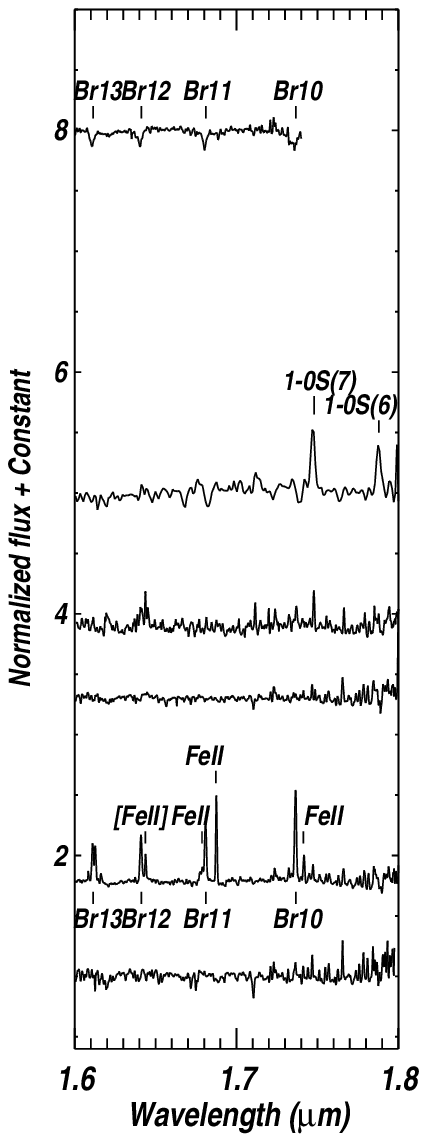}\includegraphics[bb=14 8 235 337,scale=0.65]{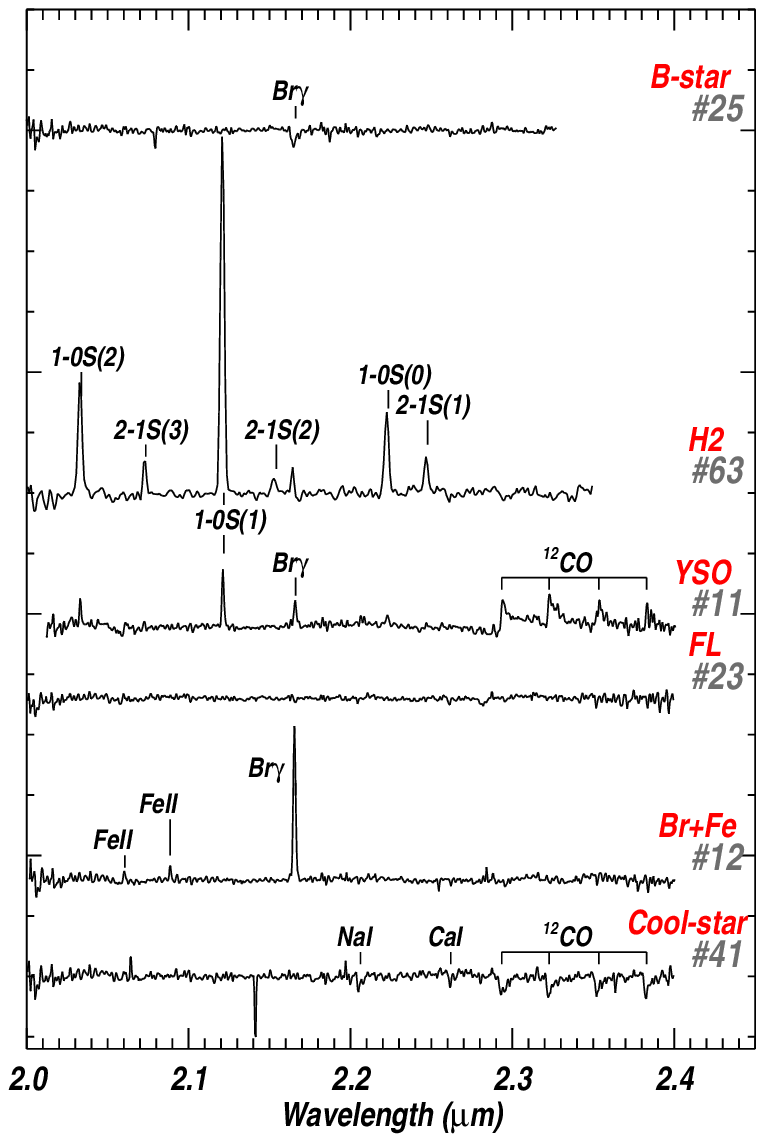}
}
\caption{Sample of representative spectra in NGC~7538.  B-star: B-type star, H2: dominant H$_2$ lines, YSO: Young Stellar Object, FL: featureless spectrum, Br+Fe: Brackett and FeII emission lines,  Cool-St: cool star. }
\label{sample_spectra}
\end{center}
\end{figure}

We present the higher-resolution spectra of the selected high-mass star candidates in Fig.~\ref{fig_spectra} and Fig.~\ref{fig_spectra_4}. The lower resolution spectra are presented in Fig.~\ref{fig_spectra_lr}.\\    
The inspection of the NIR spectra of the  21 objects observed in NGC~7538 suggests a classification into the six major groups exemplified in Fig.~\ref{sample_spectra}.

\paragraph{ Early-type stars:}
Only 3 objects within our candidate sample exhibit atmospheric spectral features in absorption, but  none of them show traces of Helium. This evidences that sources \#35, \#25 and  \#17 are B-type stars. Considering the  measured  Br-11 and Br$\gamma$ EWs, we classify the early-type stars following the  schemes described in \cite{Hanson_96} and \cite{Hanson_98}. The obtained spectral types range between B1.5 and B9V. 

IRS~6 and IRS~5 are very bright sources in the field of NGC~7538 that we did not include in our NIR spectroscopic follow-up. These objects were already classified by \cite{Russeil_07} according to the spectral features detected in their blue spectra. However, the strength of the HeI and HeII lines on these published spectra indicate that, opposite to what the authors state in the article, IRS~6 is of earlier spectral type than IRS~5. To investigate this, we performed spectroscopic observations in September 2009 with ISIS at the William Herschel Telescope, confirming that the spectral classification of \cite{Russeil_07} is swapped for these two particular objects.  This revised classification solves the disparity in the estimates of the cluster distance  derived from each star, as it will be later discussed in Sect.~\ref{sect_dist}.\\

\paragraph{ H$_2$ emission sources:}
NIR ro-vibrational emission from H$_2$ is detected toward source \#63.  This source is  located in the vicinity of the active IRS~1-3 region and amidst the strong extended H$_2$ emission reported by \cite{Davis_98} and \cite{Bloomer_98} and attributed to IRS~1/2 \citep{Kraus_06}.  The molecular hydrogen emission has indeed a thermal origin, pointing to the  presence of shocks produced by outflows.   \\
The spectra of  sources  \#60, \#11 and \#2 also show traces of H$_2$~1--0S(1) emission. However, this spectral feature has a nebular origin for the earlier case or it is not a dominant feature for the other two.

\paragraph{ CO bandhead emission sources:}
Objects \#6 and \#14 show CO first-overtone emission in their K-band spectra. These CO bandheads are emitted by neutral material with temperatures in the range 2000-5000 K and densities  of $\sim$10$^{10}$\,cm$^{-3}$ \citep{Scoville_79}.  CO bandhead and Br$\gamma$ emission are spectral features commonly displayed by YSOs \citep{Bik_06}. Surprisingly, their $H$-band spectra  show the presence of both nebular emission and Brackett features in absorption. These two candidate YSOs are fairly isolated: \#6 lies in the centroid of a trunk-like structure  at the rim of the Sh~2-158 \HII{} region in the IRAC maps, whereas  \#14 belongs to the  embedded cluster detected at the location of NGC~7538\,S  or region ({\sc v}).
Sources \#11 and \#2, located respectively  13$\arcsec$ and 2$\arcsec$ away from IRS~4, show Br$\gamma$ and CO first-overtone emission in their spectra. Contamination by IRS~4, particularly in the spectrum of source \#2, must be considered.

\paragraph{ Featureless sources:}
Sources \#23, \#36, and \#18 are identified as featureless. The NIR spectra of these objects could resemble those of low-mass Class~I objects in \citep{Greene&Lada_96} or even the Herbig Ae stars found by \cite{Rodgers_01}.

\paragraph{ Brackett+Fe emission sources:}
Two objects (\#26 and \#12) in the field show a peculiar spectrum with the Brackett series in emission and several FeII lines.\\
Source \#12 shows broad Br15-10  features (average FWHM of 220\,km\,s$^{-1}$) and  FeII lines  at 1.688, 1.742, 2.061 and 2.089\,$\mu$m and [FeII] line emission at 1.644 \,$\mu$m. For classical Be stars, the presence of Fe II emission in the $K$ band, while MgII is not detected, is observed in stars of spectral type B4-B7 \citep{Clark_00}. A hint of CO bandhead emission is observed in its K-band spectrum. Due  to the presence  of forbidden line emission, we classify \#12 as a Herbig Be.\\  
The CO bandhead and broad Br$\gamma$ emission are distinctive features associated with YSOs \citep{Bik_06}.  The spectrum of object \#26 (aka IRS~3) shows Br17-10, FeII lines and prominent [FeII] emission at 1.553, 1.644 and 1.677\,$\mu$m.   The ratio of [FeII] 1.553 to 1.644\,$\mu$m flux for this source ($\sim$0.33) is indicative of a high-density environment  $n_e>$10$^4$\,cm$^{-3}$  \citep{Hamann_94}. This is not surprising, since \#26 appears reddened in our CCD  by an A$_{\rm V}$ $\sim$27 mag. Source \#26 appears therefore deeply buried and located at only  0.2 pc in projected distance from  the high mass protostars IRS~1/2 \citep{Bloomer_98}. The presence of [FeII] lines toward \#26, together with the detection of H$_2$ emission in the same spectra,  suggest the presence of dissociative J-type shocks, tracing flow activity toward \#26 and confirming the young nature of this source.
Source \#26 has an associated \HII{} region studied in radio recombination lines and detected continuum counterparts at 3.6 and 2\,cm \citep{Sewilo_04}. These author's inferred   Lyman continuum photon rate corresponds to that of an early B-type dwarf. This result is in good agreement with our photometry and the  dereddened F(HeI 2.058)/F(Br$\gamma$)$=$0.073, which indicates temperatures lower than those produced by O-type stars  \citep{Benjamin_99}. 
We classify \#26 as an \HII{} region powered by a B star with ongoing flow activity. The presence of FeII lines may indicate that although much of the emission is from an HII region, the object is still rather young and surrounded by  circumstellar material. \\
We note the strong similarity of the $H$- and $K$- band spectra of \#26 and \#12 with some LBVs at low excitation \citep[e.g.][]{Geballe_00}, showing Brackett lines  together with permitted and forbidden FeII lines in emission.
However, the low luminosity ($\leq 10^4$L$_{\sun}$) of these objects, together with the absence of MgII and SiII lines goes against an LBV classification. This should be regarded as a warning for missidentifications in LBV infrared spectroscopic searches in our Galaxy.

\paragraph{ CO bandhead absorption sources:}
Seven objects within the sample show typical spectral features of cool stars.  Our classification of these sources follows the prescriptions of  \cite{Winston_09} and \cite{Bik_10} for late-type stars. This comparative method uses primarily the MgI, NaI and CaI  atomic lines as diagnostics for the temperature and the CO first overtone absorption bands as luminosity indicator. This classification carries an uncertainty of about $\pm$1 spectral subclass.
Stars \#46 and \#4 show spectral features compatible with that of a K0III, whereas star \#60 is classified as a K2III.  
Sources \#55, \#39, \#58 and \#41  have shallower CO absorption bands than that of giants and yet deeper than dwarfs, indicating a luminosity class IV.  Therefore, these objects are considered pre-main sequence stars (PMS).  

Despite the small offsets between the nodding positions and the narrow aperture used for extracting the spectra, nebular \HII{} emission appears strongly in the spectra of sources \#58 and \#41.  These objects  are located in the line of sight of  areas where a strong gradient in the nebular emission is detected in our NIR maps. Therefore, these lines should be considered with caution. \\
A summary of the photometric and spectroscopic characterisation of NGC~7538's stellar content is compiled in Table \ref{spc_class}.\\
The spectral type classification allows the estimate of the extinction toward several sources in NGC~7538. The visual extinctions that appear in column (8) of Table~\ref{spc_class} have been calculated  from the spectral type classification and correspondent stellar colours, considering the \citep{Rieke&Lebofsky} extinction law. The intrinsic colours of O-type dwarfs were taken from \cite{Martins_06}, while those for B dwarfs  and giants were extracted from \cite{Tokunaga_00}. In the case of the PMS stars, we considered the intrinsic colours reported by  \cite{Kenyon&Hartmann}.   
Absolute magnitudes are taken from  \cite{Martins_06} in the case of the O stars.

\subsection{Infrared excess}\label{ir_excess}

\begin{figure*}[!t]
\centering
\includegraphics[scale=0.7, bb= 0 0 640 381]{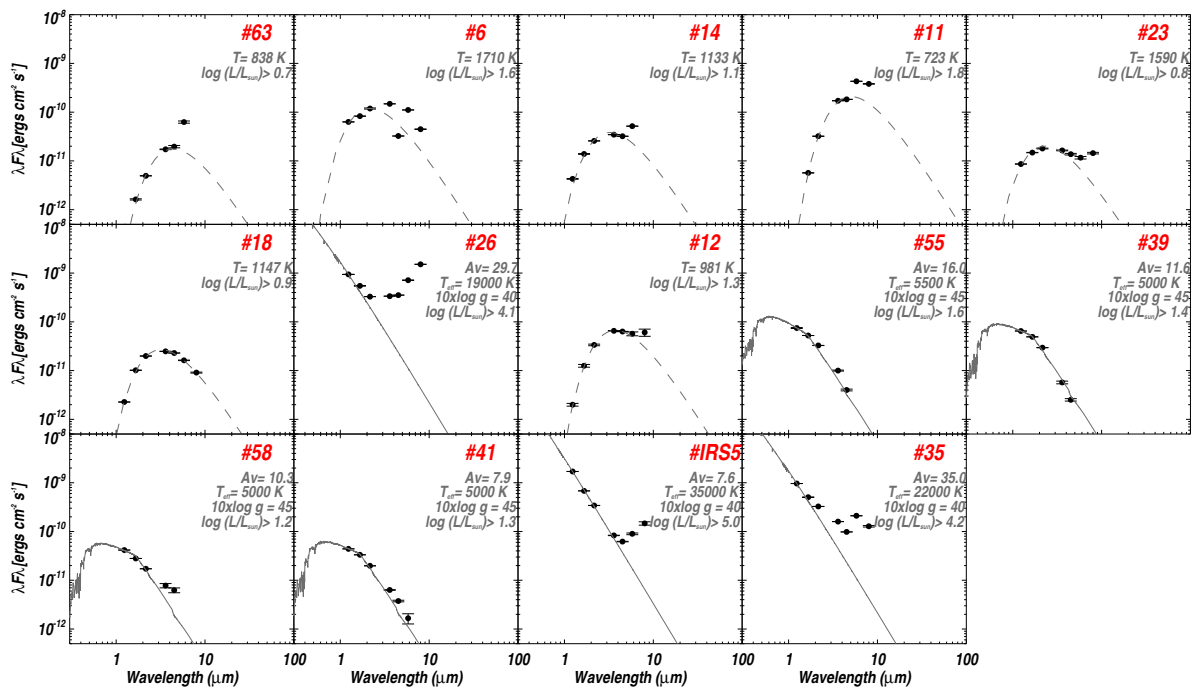}
\caption{SEDs of young sources in NGC~7538. Black dots with error bars are photometric measurements with LIRIS and IRAC. For the sources with a spectral type classification and an extinction estimate, the SED has been dereddened. Solid gray lines represent the Kurucz model fit of the respective  spectral type classification, indicated by the T$_{eff}$ and $10 \times $log $g$ labels. The area below the model in the range 0.02-1000$\mu$m corresponds to the estimated luminosity. Dashed gray lines represent the fit of a black-body distribution with the indicated temperature and luminosity.}\label{fig_yso_sed}
\end{figure*}

We have complemented our NIR photometry with  IRAC/Spitzer mid-infrared (MIR) data for this region (Chavarria et al. private communication) in order to study the infrared excess toward our follow-up sources. Fig.~\ref{fig_yso_sed} shows the spectral energy distribution (SED) of the spectroscopically identified young sources within NGC~7538.\\  
We fitted the LIRIS and IRAC flux distributions with: ({\sc i}) a Kurucz model depicted with a solid line for objects with a spectral type classification of according temperature and surface gravity; ({\sc ii})  a single black-body distribution represented by a broken line for those sources whose spectral type could not be constrained. Only in the first case, the fluxes were dereddened following the \cite{Fitzpatrick_99} parametrization. The total luminosity for each object was computed integrating the area below the respective model fit and it is considered a lower limit for objects that display an excess.

Object \#63 shows both NIR and  MIR excess, although the  8.0\,$\mu$m IRAC map appears saturated at this location. The large IRAC colour [3.6]$-$[4.5]$=$0.9 mag  hints toward  a Class~I source. \cite{Martins_09} have shown  that the H$_2$-dominated sources in the RCW~79 and RCW~120 star-forming regions have bright counterparts at 24.0\,$\mu$m. The authors conclude that this large infrared excess would be due to the existence of an envelope, implying that these H$_2$ sources would be in an early evolutionary phase.\\
However, the attribute of  massive YSO relies in the additional presence of a  strong IR excess.
\#6, \#14 and \#11 exhibit  both NIR and MIR excesses. Due to its proximity to the bright source IRS\,4, and saturation in the IRAC maps,  we can only report the NIR excess of source \#2 at sub-arcsecond spatial resolution. The photospheric features present in sources \#6 and \#14 seem indicative of  a more evolved phase, when the central star has significantly eroded the disk and photospheric features in absorption are present. This idea is supported by their IRAC colour-colour classification: Class~II in the case of \#6 and Class~I for \#14 and \#11.  However, the fits of the SEDs to  black-body distributions indicates a decreasing temperature, that could be directly related to the different spectral features observed.\\
Objects \#23, \#36 and \#18 posses  featureless NIR spectra.  \#23 and \#18 show both NIR and MIR excess, and a Class~II IRAC classification, indicating  a very young nature \citep{Greene&Lada_97}.  Source \#36 is a deeply embedded source (A$_{\rm V}$$\sim$40 mag) that displays the NIR colours
of a reddened naked photosphere without an intrinsic excess. This hypothesis is supported by the its non detection in the IRAC maps. We tentatively classify \#36 as an early-type O star whose shallow hydrogen features are not detected at the limited SNR of our NIR spectra.\\
The SED of \#26 (IRS\,3) reveals the presence of a strong MIR excess that translates into the IRAC colours of a Class~I object.\\
We have spectroscopically classified source \#12 as a Herbig Be star. The SED reveals an IR excess that rises up to 8.0\,$\mu$m. Its NIR photometry renders a reddening-free parameter Q$=$$(J$$-$$H)$$-$$1.70$$(H$$-$$K_s)$ equivalent to $-$0.29.  Herbig Ae/Be star candidates cover a range $-$1.38$<$Q$<$$-$0.22 according to \cite{Hernandez_05}, reinforcing our classification.\\
Among the identified PMS stars, only source \#58 shows a significant mid-IR excess. However, the non detection of \#55 and \#39 in the IRAC maps at 5.8 and 8.0\,$\mu$m prevents further conclusions on the IR excess of these sources.\\
Despite classified as OB stars, the SEDs of sources IRS\,5 and \#35 exhibit strong MIR excesses (i.e.~ the typical colours  of Class~I sources). In the case of \#35, this effect could be explained by the presence of a nearby object (1$\farcs$2 away), that is not resolved in the longer wavelength IRAC channels. \\

\section{Discussion}\label{discussion}

\subsection{Distance, age and mass}\label{sect_dist}

An initial estimate of the distance based on   spectro-photometry by \cite{Blitz} situated NGC~7538 at an heliocentric distance of 2.8$\pm$0.9\,kpc. 
Meanwhile, kinematical estimates  delivered a distance for the same region as far as 3.5 kpc \citep{Israel_73}.
\cite{Russeil_07} determined the spectral type classification  of two stars in the \HII{} region: IRS~6 (O9V), and IRS~5 (O3V), inferring spectro-photometric distances of 1.6$\pm$0.17 and 4.24$\pm$0.29\,kpc, respectively. They estimated an average  distance of  2.27$\pm$0.15\,kpc  for the entire Sh~2-158 region.
The most recent estimate of the distance to NGC~7538 has been obtained  measuring trigonometric parallaxes of methanol masers, regularly associated with high-mass star forming regions. This study yielded a distance of 2.65$^{+0.12}_{-0.11}$\,kpc \cite{Moscadelli}.\\ 
Our revised optical classification of the O-type sources IRS~6 and IRS~5 described in Sect. \ref{spect_class} renders a new distance estimate to NGC~7538. Considering a classification O3V for the IRS~6 object, we derive a spectro-photometric distance of 2.99$\pm$0.5\,kpc to the cluster powering the \HII{} region. Likewise, the classification of IRS~5 yields a value of 2.39$\pm$0.4\,kpc. We report a final spectro-photometric distance of 2.7$\pm$0.5\,kpc to this region.

\begin{figure}
\begin{center}
\resizebox{\hsize}{!}{
\includegraphics[bb= 24 0 248 200]{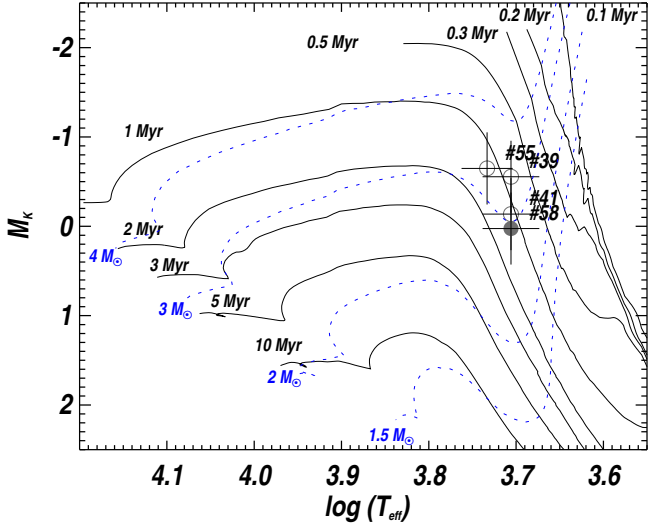}
}
\caption{Extinction corrected Hertzsprung-Russell diagram of the identified PMS stars within NGC~7538. The open symbols correspond to PMS stars without an IR excess detected in their LIRIS and IRAC photometry, while the filled ones represent those with an IR excess detected. The solid lines represent the pre-main-sequence isochrones, while the dotted lines correspond to evolutionary tracks by \cite{DaRio}.}\label{Age_Mass_pms}
\end{center}
\end{figure}

Despite the large extinction variations in the region that hampered a robust age determination, \cite{Balog_04} found an older generation of stars ($\sim$\,4 Myr) and a younger population of faint stars closer to $\sim$\,1 Myr.  A  fraction  of 30\% of young stars in the region around NGC~7538 was reported to exhibit a near-IR excess.
An upper limit limit to the age of the cluster can be established from the most massive classified star member. According to \cite{Schaerer_97}, a 60\,M$_{\sun}$ star (corresponding to a O3V star) has a main-sequence lifetime $\sim$2.2\,Myr before it starts the giant phase. Assuming a coeval star-formation event,  the oldest stellar content in NGC~7538 must be, therefore, younger than 2.2\,Myr. \\
At the other end of the mass spectrum, we can use the information derived from the PMS stars to constrain the age of the low-mass population around NGC~7538. To construct a Hertzsprung-Russell diagram (HRD), we use the spectral type to temperature conversion from \cite{Kenyon&Hartmann}, including the overestimates reported by \cite{Cohen_79} as temperature error. The absolute $K$ magnitudes are then calculated from the dereddened apparent brightness and assuming a distance of 2.7\,kpc. The HRD of the PMS stars identified in NGC~7538 is shown in Fig.~\ref{Age_Mass_pms}.  We have depicted in the figure several pre-main-sequence evolutionary tracks and isochrones taken from \cite{DaRio}. The mass of these PMS stars is estimated in the range 2-4 M$_{\sun}$.  The comparison with the PMS isochrones  yields an age range between 0.5 and 2\,Myr for the PMS identified population. We conclude, thus, that these two estimates result in an age  range between 0.5 and 2.2\,Myr for the powering cluster. \\
Clearly, other massive young stellar populations are also present at several locations around NGC~7538. They range from stars for which we observe photospheric spectral features, are detectable in the optical, but also exhibit  very strong IR excess (IRS~5 at region {\sc(ii)}) to objects whose emission is dominated by the presence of a circumstellar disk (region {\sc(iii)}), and finally to nascent stars that present on-going outflow activity (IRS~1/2 and IRS~9 in regions {\sc(iv)} and {\sc(vi)}, respectively).\\ 
These results support the previous idea of  a star-forming  sequence in the northwest-southeast direction that may be smoother than previously considered. The quantitative analysis of the individual ages of these populations is beyond the scope of this paper.

Assuming a Salpeter initial mass function (IMF), and extrapolating it down to 0.8 M$_{\sun}$, we can calculate the stellar mass of a cluster. Yet, this estimate is only meaningful for a  coeval stellar population. This should be the case for the powering cluster of the \HII{} region, with IRS~6 as the ionizing star.  Normalising the stellar mass distribution  by the detection of one O3V star with a mass uncertainty between 47 and  64\,M$_{\sun}$, we obtain a total mass $\sim$1.7$\times$10$^3$\,M$_{\sun}$ for the cluster.

\subsection{Spatial distribution}
The second most massive identified star is IRS~5 and, although is located amidst the ionised emission, its infrared excess and proximity to the rim of the molecular cloud suggest that it may not be a member of the central cluster, and region {\sc (ii)} in fact comprises a younger generation of stars. Among the spectroscopically identified B-type stars,  only object \#25 is located amidst  the ionised emission, while sources \#17 and \#35 are farther away. In the case of \#35, this source is located in the vicinity of the young region IRS~1/2 or {\sc (iv)} and exhibits an infrared excess typical of  Class~0/I objects.\\
In fact, our CMD  on the left panel of Fig.~\ref{CMD_CCD} reveals a deficit of early B-type candidates for sources located within a radius of 30$\arcsec$ from the O3V ionizing star IRS~6 (objects marked by squares). To further study this sparseness of early B-type stars associated to the cluster, we identify in the CMD B-type candidates of compatible $K_s$-band brightness and $H-K$$\sim$0.4. These objects appear indicated by diamonds in the left panel of Fig.~\ref{CMD_CCD}. Analogously to our previous analysis of spectroscopically identified B-type stars, a few of the B-type candidates show a mid-IR excess (Chavarr{\'{i}}a et al., private communication) suggesting they belong to younger pockets of star formation. Other B-type candidates  exhibit only photospheric emission and they are located to the east and shouth of IRS~6, spreading  beyond the boundary of the \HII{} region, possibly due to a mass segregation effect. 

We sketch a possible picture in which  the cavity created by the ionizing stars in  the molecular cloud is open in the observer's direction and part of its stellar population is located in the foreground of the neutral molecular cloud, appearing almost aligned with other embedded younger star-forming regions. This hypothesis is supported by the work of \cite{Balog_04} who found that the stellar density peak of this region at NIR wavelengths is located at the rim of the \HII{} region. \\
This projection effect is particularly important to bear in mind when isolating the stellar populations of the different pockets of star-formation that may be triggered by the expansion of \HII{} regions.

\section{Summary and conclusions}\label{conclusions}

We  have reported the spectro-photometric study of a sample of  candidate high-mass stars in NGC~7538.  We have used the LIRIS instrument to obtain $JHK_s$ photometry and multi-object and long-slit spectroscopy in the $H$ and $K$ bands at low and medium resolution. Complementary IRAC photometry of the sample has aided in the investigation of infrared excesses.

Besides the optically known O stars IRS\,6 and IRS\,5, whose spectral type classification we have revised,  we have  pinpointed a population of B-type stars located either at or beyond the rim of the \HII{} region. In view of this, we confirm that IRS\,6 is  the main ionizing source of Sh~2-158.   Considering the non detection of spectral features and the lack of infrared excess, we suspect that object \#36 is a deeply embedded early O-type star in the vicinity of IRS\,1-3. However, higher signal-to-noise-ratio observations are needed to ascertain this hypothesis.\\
We have spectroscopically identified four G and K stars still contracting to the main sequence, and therefore labelled them as PMS. These stars are considered to be the precursors to A and F stars.  \\
The observations of IRS\,3, aka \#26, are consistent with its previous classification as a B-type candidate that has already started powering a compact \HII{} region in a high-density environment $n_e>$10$^4$\,cm$^{-3}$ near the protostars IRS\,1-2. The outflow activity traced by the  [FeII] and H$_2$ molecular emission also support the young nature of this object.\\
Five of the  observed sources  have been classified as YSOs, being their spectra dominated either by H$_2$ or CO features. Two stars display featureless NIR spectra and a prominent infrared excess, while object \#12 has been identified as a Herbig Be star.  In general, all these YSOs are considered to be the precursors of OB stars. 

We establish a discrimination between an older population of naked stellar photospheres and various regions of more recent star formation around NGC~7538. However, the former population overspreads spatially beyond the \HII{} region likely due to a projection effect that difficults the isolation of the population associated to the individual pockets of  star formation. We find that most of the YSOs are concentrated either at the southern rim of the \HII{} region or in the area around the infrared sources IRS\,1-3. This, together with the known on-going formation of the protostars in the field,  reinforces the idea of a northwest-southeast star-forming sequence. However, the study of the detailed sequence of events and causality between them can not be addressed with the presented data.  

We have estimated a distance of 2.7$\pm$0.5\,kpc, an age range 0.5$-$2.2\,Myr,  and a total mass  $\sim$1.7$\times$10$^3$\,M$_{\sun}$  for the evolved cluster content within NGC~7538.

Due to the  variety of NIR spectral features shown by young objects in this field,  NGC~7538 is an ideal region to study   NIR spectral features of YSO for a large range of masses  and evolutionary phases within a determined environment.  

\begin{sidewaystable*}[!ht]
\vspace{15cm}
\caption{Coordinates, LIRIS photometry and spectral classification of the selected objects within NGC~7538. YSO: young stellar object; FL: featureless spectrum; PMS: pre-main-sequence. 
}
\begin{center}
\begin{tabular}{c l l l c c c c c c c}
\hline\hline
Name &  ID & R.A.\,(J2000) &    DEC\,(J2000)   &  $J$ &  $H$ &  $K_{s}$ &   A$_{\rm V}$(mag) &  Class. & ST  &   Notes \\
     &     &      &      &      &      &      &      &      &      &      \\
\hline
 IRS\,6 &  &  23 13 34.3 & +61 30 14.3 &  8.71 $\pm$ 0.02$^{\dag}$ & 8.21 $\pm$ 0.04$^{\dag}$ & 8.00 $\pm$ 0.01$^{\dag}$ & 5.4$\pm$0.4&        &   O3$\pm$1V & O9$\pm$1V$^{\ddag}$ \\
 IRS\,5 &  & 23 13 30.3 & +61 30 11.6 & 10.38 $\pm$ 0.03$^{\dag}$ & 9.82 $\pm$ 0.04$^{\dag}$ & 9.43 $\pm$ 0.04$^{\dag}$ & 7.6$\pm$0.6&        &   O9$\pm$1V & O3$\pm$1V$^{\ddag}$ \\
\hline
      & \#35  & 23  13  47.336   & +61  27  53.74 & 18.18 $\pm$ 0.03 & 14.62 $\pm$ 0.01 & 12.55 $\pm$ 0.01 &35.0$\pm$0.2&           &   B1.5-B3V       &   \\
      & \#25  & 23  13  43.820   & +61  29  28.37 & 13.41 $\pm$ 0.01 & 12.33 $\pm$ 0.01 & 11.83 $\pm$ 0.02 &9.3$\pm$0.3&           &   B2-B7V     &   \\
      & \#17  & 23  13  31.933   & +61  27  47.40 & 13.83 $\pm$ 0.01 & 12.38 $\pm$ 0.01 & 11.59 $\pm$ 0.01 &13.6$\pm$0.2&           &   B5-B9V     &   \\
      & \#63  & 23  13  45.058   & +61  27  53.88 & 18.13 $\pm$ 0.05 & 15.14 $\pm$ 0.03 & 13.18 $\pm$ 0.01 &  --       & YSO       &     --       &   H$_2$   \\
      & \#6   & 23  13  43.498   & +61  29  37.31 & 11.97 $\pm$ 0.01 & 10.87 $\pm$ 0.01 &  9.73 $\pm$ 0.03 &  --       & YSO       &     --       &  neb \\
      & \#14  & 23  13  44.538   & +61  27  15.19 & 14.89 $\pm$ 0.02 & 12.81 $\pm$ 0.02 & 11.39 $\pm$ 0.01 &  --       & YSO       &     --       &  neb, Br-abs  \\
      & \#11  & 23  13  30.688   & +61  29  07.47 & 17.50 $\pm$ 0.04 & 13.78 $\pm$ 0.01 & 11.15 $\pm$ 0.01 &  --       & YSO       &     --       &  H$_2$, neb  \\
 IRS\,4 & \#2   & 23  13  32.219   & +61  29  07.32 & 11.40 $\pm$ 0.01 & 10.43 $\pm$ 0.05 &  9.43 $\pm$ 0.09 &  --       & YSO       &     --       &  neb \\
      & \#23  & 23  13  34.240   & +61  29  09.62 & 14.13 $\pm$ 0.01 & 12.74 $\pm$ 0.01 & 11.78 $\pm$ 0.02 &  --       & FL        &     --       &   \\
      & \#36  & 23  13  45.380   & +61  28  04.19 & 20.09 $\pm$ 0.14 & 15.25 $\pm$ 0.03 & 12.61 $\pm$ 0.01 &           & FL        &     --       &   \\
      & \#18  & 23  13  57.268   & +61  28  15.63 & 15.58 $\pm$ 0.01 & 13.15 $\pm$ 0.01 & 11.67 $\pm$ 0.01 &  --       & FL        &     --       &    \\
 IRS\,3 & \#26  & 23  13  43.620   & +61  28  14.00 & 16.82 $\pm$ 0.04 & 13.67 $\pm$ 0.03 & 11.95 $\pm$ 0.03 &29.7$\pm$0.7& \HII{}       &     --       &  \\
      & \#12  & 23  13  33.537   & +61  28  49.80 & 15.72 $\pm$ 0.07 & 12.92 $\pm$ 0.07 & 11.10 $\pm$ 0.04 &  --       & Herbig Be &     --       &     \\
      & \#55  & 23  13  38.488   & +61  28  47.59 & 15.98 $\pm$ 0.02 & 13.98 $\pm$ 0.01 & 12.92 $\pm$ 0.02 & 16.0$\pm$0.3& PMS       &   G9$\pm$1IV       &   \\
      & \#39  & 23  13  40.766   & +61  28  28.20 & 14.98 $\pm$ 0.01 & 13.33 $\pm$ 0.01 & 12.54 $\pm$ 0.01 & 11.6$\pm$0.2& PMS       &   K1$\pm$1IV       &   \\
      & \#58  & 23  13  31.325   & +61  30  22.24 & 15.12 $\pm$ 0.01 & 13.73 $\pm$ 0.01 & 12.98 $\pm$ 0.01 & 10.3$\pm$0.2& PMS       &   K1$\pm$1IV       &  neb Br$\gamma$ \\
      & \#41  & 23  13  33.391   & +61  29  35.77 & 14.43 $\pm$ 0.01 & 13.15 $\pm$ 0.01 & 12.56 $\pm$ 0.01 &7.9$\pm$0.3&   PMS      &   K1$\pm$1IV/III      &   \\
      & \#15  & 23  13  57.238   & +61  27  47.40 & 14.95 $\pm$ 0.01 & 12.60 $\pm$ 0.01 & 11.56 $\pm$ 0.01 &16.5$\pm$0.2&           &   G8$\pm$1III    &   \\
      & \#46  & 23  13  34.218   & +61  30  17.64 & 14.18 $\pm$ 0.07 & 13.03 $\pm$ 0.09 & 12.66 $\pm$ 0.08 &4.9$\pm$1.4&           &   KO$\pm$1III    &   \\
      & \#60  & 23  13  44.280   & +61  28  10.10 & 15.68 $\pm$ 0.01 & 13.91 $\pm$ 0.01 & 13.05 $\pm$ 0.03 &11.9$\pm$0.6&           &   K2$\pm$1III      &   H$_2$  \\
\hline
 IRS\,7 & \#4   & 23  13  36.801   & +61  30  39.59  & 8.44 $\pm$ 0.04$^{\dag}$ & 7.63 $\pm$ 0.04$^{\dag}$ & 7.37 $\pm$ 0.01$^{\dag}$ &2.6$\pm$0.6&         &    K0$\pm$1III       &  \\
\hline
\end{tabular}
\label{spc_class}
\end{center}
$^{\dag}$ Photometry from the 2MASS Point Source Catalog.\\
$^{\ddag}$ Classification by \cite{Russeil_07} \\
\end{sidewaystable*}

\begin{acknowledgements}
We thank the referee for valuable comments that helped improving the contents of this paper. We  would also like to thank Dr. Bel\'en L\'opez for her inputs to the initial manuscript.\\
EP is funded by the Spanish MICINN under the Consolider-Ingenio 2010 Program grant CSD2006-00070: First Science with the GTC  (http://www.iac.es/consolider-ingenio-gtc). This work has been partly supported by grants AYA2007-67456-C02/AYA2008-06166-C03.\\
This publication makes use of data products from the Two Micron All Sky Survey, which is a joint project of the University of Massachusetts and the Infrared Processing and Analysis Center/California Institute of Technology, funded by the National Aeronautics and Space Administration and the National Science Foundation.
\end{acknowledgements}

\bibliographystyle{aa}
\bibliography{3294}

\end{document}